\documentclass[sigconf]{acmart} 
\acmSubmissionID{xxx}
\def\BibTeX{{\rm B\kern-.05em{\sc i\kern-.025em b}\kern-.08em
    T\kern-.1667em\lower.7ex\hbox{E}\kern-.125emX}}

\usepackage{booktabs} 
\usepackage{hyperref} 
\usepackage[most]{tcolorbox}

\acmDOI{}

\acmISBN{978-3-89318-099-8}

\usepackage{amsmath}
\usepackage{mathtools}
\usepackage{graphicx} 
\usepackage{url}
\usepackage{enumitem} 

\usepackage{stmaryrd} 

\usepackage{upgreek} 
\usepackage{bbding} 

\usepackage{listings} 
\usepackage{tabularx} 
\usepackage{booktabs} 
\usepackage{multirow}
\usepackage{hhline} 
\usepackage{diagbox}
\usepackage{pgfplotstable} 

\usepackage{xcolor,colortbl}    

\usepackage{tikz} 
\usetikzlibrary{matrix}
\usetikzlibrary{calc}
\usetikzlibrary{math}
\usetikzlibrary{arrows.meta,positioning}
\usetikzlibrary{intersections, pgfplots.fillbetween}

\usepackage{easybmat} 

\usepackage{adjustbox}
\def\eox{\unskip\kern 10pt{\unitlength1pt\linethickness{.4pt}$\diamondsuit${}}} 

\usepackage{rotating}		

\usepackage{xspace} 

\usepackage{subcaption} 
\newcommand{\hide}[1]{}

\usepackage[ruled,noend,linesnumbered]{algorithm2e} 

\usepackage{setspace} 

\DontPrintSemicolon     
\SetNlSty{}{}{}                
\SetAlgoInsideSkip{smallskip}   

\SetAlFnt{\small}			
\SetAlCapFnt{\small}		
\SetAlCapNameFnt{\small}
\usepackage{array}
\usepackage{makecell}
\usepackage[utf8]{inputenc}

\usepackage{aliascnt}  	
\usepackage{array}
\usepackage{makecell}
\usepackage[utf8]{inputenc}

\AtEndPreamble{%
    \hypersetup{colorlinks,
      linkcolor=purple,
      citecolor=blue,
      urlcolor=ACMDarkBlue,
      filecolor=ACMDarkBlue}}

\usepackage{tcolorbox}		
\tcbuselibrary{breakable,skins}		

\tcbset{examplestyle/.style={
		enhanced jigsaw,	
		colback=blue!08,	
		colframe=blue!08,	
		arc=2mm,
		boxrule=0pt,		
		left=1mm,
		right=1mm,
		left skip= 0mm,  
		right skip= 0mm, 
		top=1mm,		
		bottom=1mm,		
		breakable,		
		parbox = false,		
		before={\par\pagebreak[0]\vspace{1mm}\parindent=0pt},		
		after={\par\pagebreak[0]\vspace{1mm}\parindent=0pt},				
		bottomrule = 0mm,
		boxsep = 0mm,					
		topsep at break=0pt,			
		bottomsep at break=0pt,			
		pad at break=0mm,
		pad before break=1mm,		
		pad after break=1mm,		
		bottomrule at break=0mm,
		toprule at break=0mm,		
		}}

\usepackage{footnote}

\tcolorboxenvironment{example}{examplestyle}

\makeatletter
\DeclareRobustCommand*\uell{\mathpalette\@uell\relax}
\newcommand*\@uell[2]{
  \setbox0=\hbox{$#1\ell$}
  \setbox1=\hbox{\rotatebox{10}{$#1\ell$}}
  \dimen0=\wd0 \advance\dimen0 by -\wd1 \divide\dimen0 by 2
  \mathord{\lower 0.1ex \hbox{\kern\dimen0\unhbox1\kern\dimen0}}
}
\makeatother

\usepackage{marginnote}

\renewcommand{\epsilon}{\varepsilon} 

\usepackage{scalerel}
\usepackage{tikz}
\usetikzlibrary{svg.path}

\definecolor{orcidlogocol}{HTML}{A6CE39}
\tikzset{
  orcidlogo/.pic={
    \fill[orcidlogocol] svg{M256,128c0,70.7-57.3,128-128,128C57.3,256,0,198.7,0,128C0,57.3,57.3,0,128,0C198.7,0,256,57.3,256,128z};
    \fill[white] svg{M86.3,186.2H70.9V79.1h15.4v48.4V186.2z}
                 svg{M108.9,79.1h41.6c39.6,0,57,28.3,57,53.6c0,27.5-21.5,53.6-56.8,53.6h-41.8V79.1z M124.3,172.4h24.5c34.9,0,42.9-26.5,42.9-39.7c0-21.5-13.7-39.7-43.7-39.7h-23.7V172.4z}
                 svg{M88.7,56.8c0,5.5-4.5,10.1-10.1,10.1c-5.6,0-10.1-4.6-10.1-10.1c0-5.6,4.5-10.1,10.1-10.1C84.2,46.7,88.7,51.3,88.7,56.8z};
  }
}

\RequirePackage{etoolbox}
\DeclareRobustCommand\orcidicon[1]{\href{https://orcid.org/#1}{\mbox{\scalerel*{
\begin{tikzpicture}[yscale=-1,transform shape]
\pic{orcidlogo};
\end{tikzpicture}
}{|}}}}

\usepackage{array}
\usepackage{makecell}
\usepackage[utf8]{inputenc}

\newcommand{\baseline}{\textsc{Unstructured Semantic Search}}
\newcommand{\oursearch}{\textsc{Structured Semantic Search}}

\setcopyright{none}
\renewcommand\footnotetextcopyrightpermission[1]{}

\begin{document}
\title{Diversed Model Discovery via Structured Table Discovery}

\author{Zhengyuan Dong}
\affiliation{
  \institution{University of Waterloo}
  \city{Waterloo}
  \state{ON, Canada}
}
\email{zhengyuan.dong@uwaterloo.ca}

\author{Ren\'ee J. Miller}
\affiliation{%
  \institution{University Waterloo}
  \city{Waterloo}
  \state{ON, Canada}
  }
\email{rjmiller@uwaterloo.ca}

\renewcommand{\shortauthors}{}

\begin{abstract}
Model cards describe the behavior of models through a mixture of textual descriptions and structured artifacts, including performance, configuration, and dataset tables.
Existing model search systems rely predominantly on semantic similarity over text, which can produce homogeneous result sets and limit users’ ability to explore alternatives and reason about trade-offs.
We argue that model search is inherently comparative: users want models that are aligned at the task level yet differentiated in measurable ways.
We hypothesize that this balance requires retrieval over condensed, high-quality evidence rather than verbose descriptions, and much of that evidence is concentrated in structured tables.
We present
\oursearch\ 
, a table-driven model search framework built on the curated \textsc{ModelTables} benchmark.
Given a query, 
\oursearch\ 
combines a semantic baseline for task alignment with a structure-aware pipeline that discovers query-related model-card tables using table discovery operators such as unionability, joinability, and keyword search.
Retrieved tables are mapped back to model cards under a controlled top-$k$ budget, enabling fair comparison between text-based and table-based retrieval.
Beyond retrieval, 
\oursearch\ 
adapts table integration to the model-table domain through orientation-aware integration, producing compact integrated views of tables from partially overlapping and sometimes transposed evidence tables.
For evaluation, we introduce a nugget-based, auditable protocol that extracts compact evidence items from model cards, matches queries to condition- or intent-specific nuggets, and measures evidence coverage and diversity over retrieved model-card candidate sets.
This protocol also provides a scalable path toward approximate, evidence-based labeling in dynamic model lakes.
Experiments on a 597 model-recommendation query set show improved nugget coverage for the structure-aware pipeline compared to semantic baselines.
\end{abstract}

%
%



\maketitle

\section{Introduction}

\paragraph{\bf Model search is not document retrieval.}
Model lakes
~\cite{DBLP:conf/edbt/PalBM25}
have emerged as a central infrastructure for organizing and 
sharing machine learning models.
Each model is accompanied by a model card describing training data, evaluation results, and intended usage~\cite{modelcardpaper}.

Existing model search systems, such as HuggingFace~\cite{huggingface-semantic-search}, Modelscope~\cite{modelscope}, ModelDB~\cite{mcdougal2017twenty}, TensorFlow Hub\footnote{https://www.tensorflow.org/hub}, PyTorch Hub\footnote{https://pytorch.org/hub/}, DLHub\footnote{Deep Learning Hub. https://dlhub.app/}, treat model cards as unstructured documents. 
These systems commonly rely on keyword search, metadata filters, faceted search, or semantic retrieval over model descriptions and model-card text.
While these mechanisms are effective for finding individually relevant models, they provide limited support for constructing comparison-oriented candidate sets of models.
However, model search in model lakes often requires more than retrieving individually relevant models~\cite{ma2025huggingr,li2023metadata}.
Users may want a set of task-aligned models that also differ in meaningful ways, such as architecture, training corpus, evaluation benchmarks, model variants, or performance trade-offs.
This creates a need for diverse model discovery~\cite{agrawal2009diversifying}: the result set should remain relevant to the query while exposing non-redundant alternatives for comparison.
This need aligns with the broader information-retrieval view that useful search results should balance relevance with diversity and coverage of user intents

\paragraph{\bf The tension between task alignment and diversity.}
This observation highlights a fundamental tension in model search.
On one hand, retrieved models must be aligned at the task or topic level to remain relevant.
On the other hand, users expect diversity in the results, enabling comparison and informed decision-making~\cite{ziegler2005improving}.
Pure semantic similarity optimizes for textual proximity and therefore tends to collapse results around dominant model families (for example, collections of related models developed by an organization)
limiting exposure to alternative approaches.
This effect is amplified by shared writing templates and reporting conventions: models developed by the same authors or within the same model family often exhibit highly similar narrative descriptions, even when their empirical behaviors differ~\cite{DBLP:journals/corr/abs-2512-16106}.
This tension suggests that model search should not be optimized for maximal similarity, but for controlled differentiation under task alignment.
Achieving this balance requires retrieval signals that go beyond surface-level text similarity and are less sensitive to representational and stylistic bias.
\paragraph{\bf Condensed evidence in model cards.}
Model cards contain a mixture of narrative text and structured artifacts~\cite{modelcardpaper}.
While textual descriptions provide contextual information, they are often verbose, heterogeneous, and shaped by authorial style and templating practices, making direct comparison difficult
~\cite{model-card-example}.
In contrast, structured tables, including performance summaries, benchmark results, and configuration listings, concentrate high-density, decision-critical evidence with limited stylistic freedom~\cite{kim2012scientific}.
These tables encode the core empirical claims of a model and often vary meaningfully even between closely related models~\cite{DBLP:journals/corr/abs-2512-16106}. 
By filtering out irrelevant content and normalizing how evidence is presented, tables provide a more stable basis for comparison.
This work explores how such condensed, table-grounded evidence can be leveraged to better support the inherently comparative nature of model search.

\paragraph{\bf Nugget-based evaluation.}
Model lakes evolve rapidly and user queries vary in specificity: some queries contain explicit conditions (e.g. "4-bit quantized model on X benchmark"), while others are intentionally vague (e.g. "works well on legal documents"). These characteristics make constructing a fixed gold-standard labeling impractical. To evaluate retrieval quality under these constraints, we adopt a nugget-based evaluation ~\cite{pradeep2025great} with two stages: (1) a card-to-nugget extraction step that pulls compact evidence ("nuggets") from model cards; and (2) a query-to-nugget matching, filtering, and aggregation step that maps queries to condition- or intent-specific nuggets and computes a nugget coverage score for candidate sets.

As for nugget definition, prior work varies widely (e.g., sub-questions, atomic facts, or feature-name sets). Concretely, we define  \emph{nuggets} as a set of tuples with fixed attributes  (Model, Base model, Model variant, Dataset, Metric name, Metric value).  This definition  follows the leaderboard-style atomic extraction~\cite{kardas2020axcell}.

\paragraph{\bf Contributions.}
We summarize our contributions as follows:
\begin{itemize}
\item A table-driven model discovery pipeline that complements semantic (text-based) retrieval by searching and integrating structured tables extracted from model cards.
\item A nugget-based evaluation metric and two-stage pipeline (leaderboard-derived item extraction + prompt-assisted query-to-nugget matching) that measures evidence coverage and diversity; the metric is explicitly scoped to evaluate the nuggets extracted from retrieved candidate sets (not full model-card processing) and supports approximate, evidence-based labeling in dynamic model lakes.
\item A practical integration strategy that is orientation-aware (handling tables that have been transposed) to improve comparability across retrieved evidence; from a downstream-integration perspective, the retrieved set should be visibly relevant yet diverse, and integration provides a convenient, user-facing view for side-by-side comparison.
\item An end-to-end implementation that allows inspection of retrieved tables and integration views, together with an adapted model-recommendation query set derived from paper-recommendation data; experiments using this query set show improved nugget coverage for our pipeline compared to semantic baselines.
\end{itemize}

Our work is evaluated over 60K models from HuggingFace~\cite{DBLP:journals/corr/abs-2512-16106} and the system will be demonstrated at the workshop.\footnote{All codes, prompts, data, and outputs are included in our github: \url{https://github.com/RJMillerLab/ModelSearch}.}  

\section{Related Work}
\label{sec:related-work}

\subsection{Model Lake}
Model lakes have recently emerged as a research topic for managing large collections of heterogeneous machine learning models and their associated artifacts, as envisioned by Pal et al.~\cite{DBLP:conf/edbt/PalBM25}. The model-lake literature spans tasks such as model attribution and provenance tracking~\cite{mei2022provenancemanagement, mu2023model, wang2024mitigatingdownstreammodelrisks}, model versioning and lineage analysis~\cite{DBLP:journals/tods/LeventidisRGMR23, DBLP:journals/pvldb/ShragaM23}, model search and retrieval~\cite{lu2023content, li2024model}, benchmarking and reporting~\cite{modelcardpaper, liang2024whats}, and documentation generation~\cite{DBLP:conf/naacl/0004LJD24}. Model cards are a central source of that evidence: they record model details, intended use, training data, evaluation results, and limitations~\cite{modelcardpaper}. Yet later studies show that such documentation is often incomplete, inconsistent, or hard to compare across models~\cite{liang2024whats}, which is why prior work has explored metadata representations for queryable repositories~\cite{li2023metadata}, task and model embeddings for retrieval~\cite{9008292}, content-based model search~\cite{lu2023content}, graph-based model selection~\cite{li2024model}, and LLM-based orchestration over model descriptions~\cite{shen2023hugginggpt}. More recent work extends the selection side of this space by ranking unseen models on unseen datasets from leaderboard-style tuples~\cite{cai2026modellens}. Together, these works frame model search as a structured model-selection problem driven by heterogeneous documentation rather than text similarity alone.

\subsection{Data Discovery}
Data discovery studies how users find useful datasets and tables in large, heterogeneous data lakes~\cite{DBLP:conf/icde/FernandezAKYMS18, DBLP:conf/sigmod/Fan00M23}. 
Disambiguation in data lakes has been studied to resolve homographs and make table evidence comparable across sources~\cite{DBLP:journals/tods/LeventidisRGMR23, DBLP:journals/pvldb/ShragaM23}. 
Annotation-oriented work further treats table labeling and schema-level description as a core step for making heterogeneous tables searchable~\cite{korini2022sotab}. 
For tabular data, table search aims to retrieve tables relevant to a query~\cite{DBLP:conf/edbt/ChristensenLLRM25, DBLP:conf/sigir/LeventidisCLRHM24, DBLP:journals/pvldb/Christodoulakis20}, while joinable search aims to identify tables that can be linked through shared entities or values~\cite{khatiwada2022integrating, 2023_dong_deepjoin}. 
Unionable search instead focuses on tables with compatible schemas or semantically aligned columns~\cite{DBLP:conf/sigmod/KhatiwadaSM23, 2023_hu_autotus, DBLP:journals/pacmmod/KhatiwadaFSCGMR23}. 
Unified discovery systems combine these operators in a single workflow~\cite{DBLP:journals/corr/abs-2310-02656}, and table integration completes the pipeline by aligning and combining related tables into consolidated views for downstream analysis and comparison~\cite{DBLP:conf/edbt/KhatiwadaSM26a}. 
This body of work motivates treating table search and table integration as complementary parts of one discovery pipeline when the goal is to assemble comparable evidence from fragmented tabular sources.

\subsection{Nugget Analysis and Evaluation}
Traditional retrieval metrics such as nDCG~\cite{jarvelin2002cumulated}, MAP~\cite{schutze2008introduction}, and RBP~\cite{moffat2008rank} evaluate relevance at the document level, but they do not directly measure whether a retrieved set covers the full breadth of a user's information need. 
Nugget-based evaluation addresses this limitation by decomposing answers into atomic information units in QA~\cite{voorhees1999trec, lin2007deconstructing}, while the pyramid method evaluates summarization outputs through summary content units~\cite{nenkova2004evaluating}. 
In retrieval, this coverage perspective is closely related to search result diversification, where $\alpha$-nDCG measures novelty and redundancy-aware gain~\cite{clarke2008novelty}, IA-ERR models intent-aware ranking quality~\cite{chapelle2011intent}, and Subtopic Recall measures how many distinct subtopics are covered~\cite{zhai2015beyond}. 
Recent RAG and report-generation evaluations further adopt nugget-based coverage, since missing evidence in retrieval can lead to incomplete generated answers~\cite{pradeep2024initial, samuel2026coveragebench}. 
This coverage perspective is also relevant to model search, where effective comparison requires not only retrieving relevant model documentation, but also covering complementary evidence about capabilities, benchmarks, datasets, metrics, and constraints.

\subsection{Leaderboard Generation}
Leaderboards are widely used to summarize experimental progress by organizing methods, datasets, metrics, and performance results into comparable rankings. Prior work extracts tasks, datasets, evaluation metrics, and numeric scores from machine learning papers~\cite{hou2019identification, kardas2020axcell}, and follow-up work extends this with table-centric extraction and organization~\cite{yang2022telin, kabongo2024orkg}. More recent work studies LLM-based performance tracking and benchmark construction for scientific leaderboards~\cite{csahinucc2024efficient, singh2024legobench, wu2025league}. We borrow only the tuple-oriented view from this literature: it is a convenient way to represent performance evidence, but leaderboard construction itself is not the target of this work.

\section{Methodology}
\label{sec:methodology}

\begin{figure*}
    \centering
    \includegraphics[width=\linewidth]{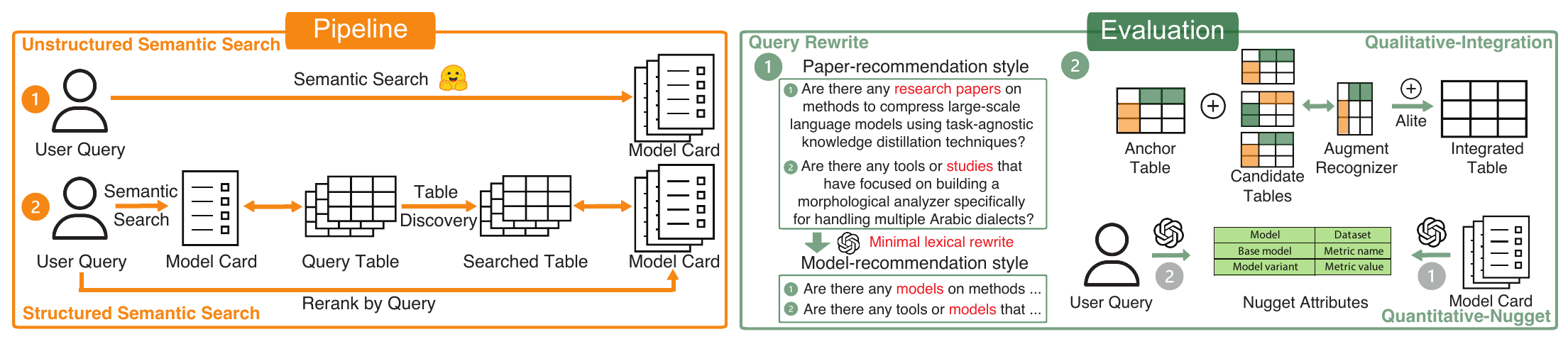}
    \caption{Overview of our table-driven model search and evaluation workflow. The pipeline augments semantic model search with table discovery and reranking, while evaluation is based on rewrite paper-style queries and measures candidate quality through table integration and nugget coverage.}
    \label{fig:pipeline}
\end{figure*}

Current deployed model search for keyword or natural language queries uses model cards~\cite{huggingface-semantic-search, hugging} . 
We will use this as our baseline (\emph{NL2Card}) that we call 
\baseline.
We also proposed a new type of model search using a table-aware candidate-generation pipeline (\emph{NL2Card2Tab2Card}) that we call 
\oursearch.
We describe each below.

\subsection{\baseline}
\label{sec:semanticSearch}

\emph{NL2Card} can be done using basic semantic search over the semi-structured model cards of a model lake.
In Figure~\ref{fig:pipeline} on the left, we depict this traditional semantic search for model cards (and their associated models) in Pipeline 1. 

Our experiments will use three implementations of semantic search:  dense, sparse, and hybrid.  
Dense retrieval is implemented with a Sentence-BERT encoder and FAISS~\cite{douze2024faiss}.
We also support sparse retrieval with Pyserini~\cite{Lin_etal_SIGIR2021_Pyserini} and a hybrid variant that retrieves an expanded sparse candidate pool before dense reranking.
The experiments report results on all three variants.

\subsection{\oursearch}
\label{sec:TableSearch}

To improve the quality and diversity of this search, we proposed leveraging the knowledge rich tables found in a model lake.  We first use \emph{NL2Card} (semantic search) to find an anchor model card, that is, the top-1 \emph{NL2Card} ranked card.  
We then use the tables associated with the anchor card in a table discovery search process described formally below.   These tables are associated with one or more models.   
Our pipeline, called 
\oursearch, uses a query-to-card-to-table-to-card workflow and is shown in Figure~\ref{fig:pipeline} Pipeline 2.   We detail each step below.

This design isolates the effect of table discovery: the semantic retrieval of an anchor model card ensures that we are finding a model associated with query task,
while table discovery expands the candidate set of models through structured evidence such as shared benchmarks, metrics, identifiers, and configuration attributes.

\subsubsection{Structure-Aware Table Discovery}

The structure-aware pipeline begins with an anchor model card selected by  \baseline.
Because table discovery requires structured evidence, the anchor step is constrained to model cards with at least one associated table.
For each anchor table, defined as a table associated with an anchor card,
\oursearch\ 
searches the model table lake using 
table discovery operators implemented in Blend~\cite{DBLP:conf/icde/EsmailoghliSMA25}.
Specifically, we use three Blend operators for
keyword search over tables (data and metadata), joinable table search, and unionable table search.

\paragraph{\bf Keyword Table Search.}
Keyword search retrieves tables containing tokens from a query set. In model tables, semantic labels and identifiers are typically concentrated in headers and first-column values (examples include benchmark task names, model, or dataset identifiers), while interior cells often contain numeric measurements and scalar values; both are informative for discovery, but play different roles. We therefore construct  keyword queries over the header and first column of an anchor table, execute Blend's value-based table keyword search operator, and rank candidate tables by matched-token frequency.

\paragraph{\bf Joinable Table Search.}
Joinable table search retrieves tables that can be join with a column 
of an anchor table~\cite{DBLP:journals/corr/ZhuNPM16}.
In model tables, joinable columns ften correspond to model names, dataset names, task names, or benchmark identifiers.
We use
the first column of an anchor table as the query column and retrieve 
joinable tables using Blend.
Tables with 
larger overlap in the join columns are ranked higher.

\paragraph{\bf Unionable Table Search.}
Unionable table search retrieves tables whose columns can be aligned with an anchor table so that their contents can be meaningfully unioned (or outer-unioned if some columns do not align)~\cite{DBLP:journals/pvldb/NargesianZPM18}.
As an example, this operator is especially useful for finding benchmark or configuration tables that report comparable attributes for different models.
We rank candidate tables by the number of distinct anchor columns that can be aligned.

\subsubsection{Mapping Tables Back to Model Cards}

Table discovery naturally returns tables, but the final retrieval task needs to return model cards.
After table discovery, we have a ranked list of tables each of which is associated with one or more model cards.
A table can be associated with more than one card if, for example, it is from a paper referenced by two or more model cards~\cite{DBLP:journals/corr/abs-2512-16106}.
For each table, we select a single model card.  To do this, we select the model card with the highest semantic retrieval similarity (using \baseline) to the query.
This table-wise top-1 selection ensures that each retrieved table contributes a single representative card, which avoids inflating the candidate set with multiple cards supported by the same table evidence.
The resulting model-card candidates (one per table) are then also ranked by their semantic query similarity and the top-$k$ selected.
Algorithm~\ref{alg:card2tab2card} illustrates the end-to-end \emph{NL2Card2Tab2Card} retrieval procedure used in our method.

\begin{algorithm}[t]
\caption{NL2Card2Tab2Card Candidate Generation}
\label{alg:card2tab2card}
\KwIn{Query $q$; model-card corpus $C$; table lake $L$; top-$k$ budget}
\KwOut{Top-$k$ model-card candidates $R$}

$A \gets \textnormal{UnstructuredSemanticSearch}(q, C)$\;
$T_{seed} \gets []$\;

\ForEach{$a \in A$}{
    $T_{seed} \gets T_{seed} \cup \textnormal{Tables}(a)$\;
}

$T_{ret} \gets []$\;

\ForEach{$Q \in T_{seed}$}{
    $T_{cand} \gets \textnormal{Discovery}(Q, L)$\;
    $T_{ret} \gets T_{ret} \cup T_{cand}$\;
}

$M \gets \emptyset$\;

\ForEach{$T \in T_{ret}$}{
    $C_T \gets \textnormal{MapTableToCards}(T)$\;
    $m^* \gets \textnormal{RerankTableCandidates}(C_T, q)$\;
    $M[T] \gets m^*$\;
}

$R_{cand} \gets \{M[T] : T \in T_{ret}\}$\;
$R \gets \textnormal{FinalRerank}(R_{cand}, q)$\;
\Return{$\textnormal{Truncate}(R, k)$}
\end{algorithm}

\section{Model Ranking Evaluation Strategy}
\label{sec:nugget-eval}

Our goal is to compare the evidence surfaced by the baseline, \baseline\  and  our new table-based search, \oursearch.  We will not consider how to achieve a static model-ranking benchmark (which to the best of our knowledge does not exist).
This matters because model lakes are continuously expanding: any fixed ground-truth annotation quickly becomes stale as new models are added.
We therefore need a comparative evaluation method that is query-aware, evidence-oriented, and stable under growth of the model lake.  We present a quantitative comparative evaluation strategy in Section~\ref{sec:quant} followed by a table-based qualitative evaluation proposal in Section~\ref{sec:qual}.

\subsection{Nugget-based Quantitative Evaluation}
\label{sec:quant}

To meet this need, we adopt a recent nugget-based strategy from information retrieval~\cite{pradeep2024initial}.
The nugget formulation lets us represent query-relevant evidence as compact, auditable units rather than as coarse document labels.
While this strategy has recently been proposed for documents, to the best of our knowledge, it has not been used for model cards.
In our setting, the same query may be satisfied by several model cards that differ only in fine-grained evidence, so the evaluation will count the evidence units explicitly.
The 
{\tt Evaluation} block on the right side of Figure~\ref{fig:pipeline} illustrates the nugget-based evaluation setting.  Before giving the formal details, we present an example.

\begin{example}
Consider the query: ``Could you recommend models that evaluate the performance decline in various
language models, like BLOOM, under 4-bit integer columnar weight-only quantization?''   
This and similar model queries refer to standard concepts like model variant ("quantization" is a model variant) or metrics (in this query, the metric name is "quantization bits" and its value is "4-bit").  We define several common concepts found in model searches and model cards.  These form the nuggets that we will extract from the model cards.  We also map the query to nuggets viewing the query as a set of constraints that the nuggests of a retrieved model card should satisfy.  
For this query, over the HuggingFace model lake we will use in our experiments (Section~\ref{sec:data}), the dense retrieval version of \baseline\
returns 10 candidate model cards, including
\textbf{inventbot/Mixtral-8x7B-Instruct-v0.1-offloading-demo},
Card-level nugget extraction yields 52 raw nugget rows in total.
For example, from the model, we extract a nugget stating the model variant is quantization, the metric quantization value is 4-bit, the metric groupsize has value 64, the metric compression has value 0, and many others.  The first nugget matches the query's nugget.
This model matches the query need.
\end{example}

\paragraph{\bf Nugget Definition}
\label{dfn:nugget}
The nugget concept is designed to correspond to atomic evidence units that are used to represent fine-grained information needs.
We adapt that idea to model search by defining each nugget as a structured tuple with a fixed set of six attributes:  
 \texttt{model}, \texttt{base model}, \texttt{model variant}, \texttt{dataset}, \texttt{metric name}, and \texttt{metric value}.
 A nugget then is a 6-tuple with one value (or null) for each of the sex attributes.  Notice some values can be null as not all models have base models and not all model cards mention the model variant.  
For a model card $c$, we denote by $\mathcal{N}(c)$ the set of nuggets extracted from $c$.
 
This fixed attribute list is important for three reasons.
First, it standardizes the evidence representation so that different model cards can be compared in the same schema.
Second, it is faithful to the structure of model cards, where these fields are commonly used to describe performance and model identity~\cite{modelcardpaper}.
Third, it makes the evaluation transparent: the same query always maps to the same kind of evidence, which is easier to inspect and trust than an open-ended free-text annotation~\cite{samuel2026coveragebench}.  

\begin{example}
Consider the model card for \textbf{luisra/Kimi-K2-Instruct-4bit},  two example nuggets extracted from this card are:  
(\textit{luisra/Kimi-K2-Instruct-4bit}, \textit{moonshotai/Kimi-K2-Instruct}, \textit{null}, \textit{null}, \textit{null}, \textit{null}) and (\textit{luisra/Kimi-K2-Instruct-4bit}, \textit{null}, \textit{null}, \textit{LiveCodeBench v6}, \textit{Pass@1}, \textit{0.537}). 
The first nugget captures the model lineage by linking the card to its base model, while the second nugget records benchmark performance on LiveCodeBench v6. 
Together, these nuggets illustrate how a single model card can contain multiple kinds of structured evidence under the same fixed schema.
\end{example}

\paragraph{\bf Nugget Extraction}

Given a model card, we extract nuggets by feeding the full card content into a prompt-based extractor that is instructed to populate our fixed nugget schema with six attributes.
The card may contain performance evidence in tables, tags, benchmark summaries, or evaluation subsections, 
so the prompt is designed to recover structured evidence from heterogeneous formatting.
We then normalize each extracted item into an instantiated nugget under the fixed schema.

The output of this stage is a nugget table containing a seventh attribute storing the model card identifier.
Because the schema is fixed and the prompt operates on the full model card, extraction only needs to be run for newly added model cards at ingestion when the model lake expands; existing nuggets do not need to be recomputed.
This makes the representation suitable for continuously growing model lakes.

\paragraph{\bf Query-to-Nugget Mapping}

The query is not itself a nugget, so we introduce an intermediate mapping from query text to nugget constraints.
The mapping identifies a subset of the nugget attributes that are relevant to the query and, when the query is specific enough, additional constraints on the attribute value of the nugget.
For vague queries, the mapping is broad and only requires attribute compatibility.
For detailed queries, the mapping includes both attributes and values, such as a benchmark name, a quantization level, or a dataset name.

We use a prompt-based method to map a query $q$ to a standardized representation $\phi(q)$ that contains the relevant nugget attributes and any associated constraints.
The query-relevant nugget set is then the subset of instantiated nuggets that match $\phi(q)$ after normalization and disambiguation.
In principle, this could be expressed as a SQL-style filter over the nugget table, but we do not rely on a literal exact matching in practice.
Many model search queries are vague or semantically under-specified, and exact field equality would miss valid evidence.
We therefore use a prompt-based filter that interprets the query intent and normalizes semantically equivalent mentions before selecting the query-relevant nuggets.
The prompt input, model output, and post-processed representation are recorded for each query so the mapping is auditable and reproducible.

\paragraph{\bf Candidate-Set Scoring}

We use a single quantity-based score to compare retrieval methods at the model card level.
For a retrieved candidate set of model cards $R_q = \{m_1, \dots, m_k\}$, we count the number of unique query-relevant nuggets covered by at least one retrieved model card.  To do this, we first take the set union of the nuggets of all model cards in the candidate set.\footnote{This is the 6-tuple nuggets without the model card id.}  The {\em Nugget Score} $\mathrm{Score}(R_q, q)$ is then the number of nuggets in this set satisfying the query constraints $\phi(q)$.

This score treats overlap carefully.
If the same nugget appears in multiple retrieved model cards, it is counted only once at the set level, preventing redundant evidence from inflating the result.
This is especially important for similarity-based retrieved sets, where highly similar model cards often contain overlapping evidence.
The score therefore directly measures how much distinct query-relevant evidence the candidate set surfaces.

In our experiments, we will compute this score for the \baseline\ methods and for the \oursearch\ methods and compare their nugget counts under the same top-$k$ budget.
Because the score is based on set coverage rather than document rank, it reflects the amount of evidence available for downstream inspection by a model lake user.

\subsection{Table-based Qualitative Evaluation}
\label{sec:qual}

In addition to the nugget-based model-card score, we 
integrate the retrieved tables and present them to the user.
The Qualitative Integration block in Figure~\ref{fig:pipeline} illustrates the table-alignment view used for manual inspection.
Retrieved tables are often partially overlapping, noisy, or transposed due to augmentation, so the integration step must be orientation-aware and iterative.
Our implementation is built on ALITE~\cite{khatiwada2022integrating}, 
a scalable approach to
integrating data lake tables that maximally integrates facts scattered
across tuples in different tables.

\paragraph{\bf Table orientation.}
To handle tables that may be transposed, we add an orientation-recognition step before integration.
For each table pair, we compare header keywords in one table against both the header row and the first column of the other table.
When the overlap pattern suggests that the two tables are semantically aligned but transposed, we transpose one table before integration.
This patch prevents direct integration from producing two poorly aligned blocks with many missing values.
Algorithm~\ref{alg:orientation_integration} summarizes this orientation-aware  integration procedure.

\begin{algorithm}[t]
\caption{Orientation-aware Integration}
\label{alg:orientation_integration}
\KwIn{Query table $Q$; retrieved tables $R=[T_1,\dots,T_n]$}
\KwOut{Integrated table $I$}

$I \gets Q$\;
\ForEach{$T \in R$}{
    $M \gets \textnormal{OverlapMatrix}(I, T)$\;
    $tr \gets (M[0,1] > 0)$ \textbf{and} $(M[0,0] = 0)$ \textbf{and} $(M[1,1] = 0)$\;
    \If{$tr$}{
        $T \gets \textnormal{Transpose}(T)$\;
    }
    $I \gets \textnormal{Integrate}(I, T)$\;
}
\Return{$I$}\;
\end{algorithm}

The resulting integrated view is the table-level counterpart to the nugget-based candidate-set score: one asks how much distinct evidence is surfaced, and the other asks whether that evidence can be organized into a coherent, comparable table.

\section{Experiments}
\label{sec:experiments}

We evaluate our proposed structure-aware pipeline against the text-only baseline on a set of almost 600 model-search queries.
We first present the model lake we use and the queries.  We then present our quantitative evaluation, which measures the number of nuggets satisfying a query returned by each search method (Section~\ref{sec:quantitative}.  
We conclude this section with two examples to better illustrate the search.

\subsection{Dataset}
\label{sec:data}

\paragraph{\bf Model Lake.}
We use the curated \textsc{ModelTables} corpus~\cite{DBLP:journals/corr/abs-2512-16106} as the  model lake.
The corpus includes over 60K model cards extracted from HuggingFace and provides a high-quality, deduplicated set of model tables associated to these model cards.  
The tables were extracted from the model cards, from code repositories referred to in the model cards, and from papers reference in the model card.  
The deduplication is important for table search: it prevents the retrieval stage from repeatedly surfacing identical tables and therefore improves the diversity and utility of the evidence returned by table discovery.
We preprocess the tables keeping only
compact tables with fewer than 200 rows and 100 columns, since the smaller tables are more likely to summarize model behavior, benchmark results, configuration attributes, or deployment constraints.
Because the same table may still be associated with multiple model cards, we keep the full table-to-card linkage during retrieval and mapping so that reranking can select one representative model card per retrieved table.
This makes the final candidate set for \oursearch more diverse than \baseline,
since it is driven by distinct table evidence rather than repeated copies of the same table.

\paragraph{\bf Query corpus}
We derive our model-search queries from LitSearch~\cite{ajith2024litsearch}, a scientific literature retrieval benchmark whose 597 queries are phrased as paper-recommendation requests.
To adapt these queries to model search, we apply a prompt-based rewrite that makes the smallest natural lexical edit while preserving the original information need.
The rewrite prompt preferentially substitutes paper-oriented terms such as {\tt paper, studies, publications, articles,} and {\tt literature} with model-oriented terms such as {\tt model, method, approach, benchmark,} or {\tt task} when appropriate, but otherwise keeps the wording and structure unchanged.
This yields a query set that retains the exploratory and comparative character of LitSearch while shifting the target from papers to models.
Importantly, we do not introduce new constraints during rewriting, so the adapted queries remain close to the original retrieval intent instead of becoming synthetic model-search prompts.
The Query Rewrite block in Figure~\ref{fig:pipeline} shows how we construct the evaluation query corpus.

\paragraph{\bf Query Type Exploration}

Query type matters for retrieval because different query intents stress different parts of the pipeline.
Prior work has proposed type-aware query taxonomies for non-factoid question answering and RAG-style evaluation, including Typed-RAG~\cite{lee2025typed}, HotpotQA~\cite{yang2018hotpotqa}, FinRAGBench-V~\cite{zhao2025finragbench}, UniDoc-Bench\cite{peng2025unidoc}, and standardized model-card question answering~\cite{toma2025answering}.
Although these schemes differ in granularity and domain, they consistently include at least evidence-seeking and comparison-oriented intents categories.
We therefore adopt the six-category non-factoid taxonomy~\cite{lee2025typed}, with the labels evidence-based, comparison, experience, reason, instruction, and debate.

\begin{figure}[!ht]
    \centering
    \includegraphics[width=\linewidth]{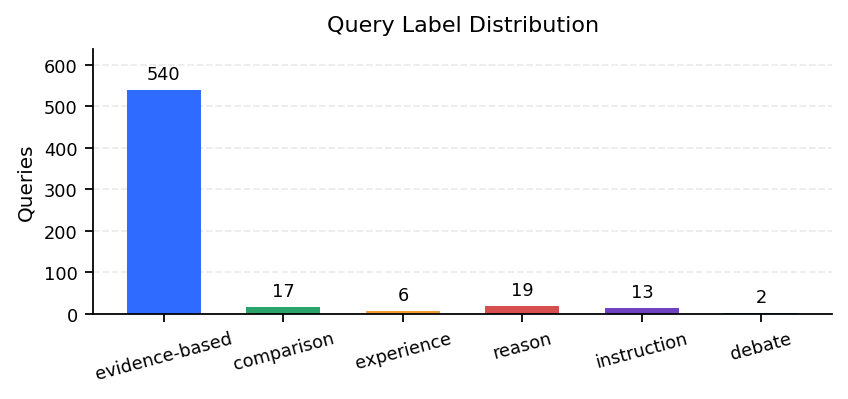}
    \caption{Distribution of query intents in the adapted LitSearch benchmark.}
    \label{fig:query_type_distribution}
\end{figure}

We label each of the 597 transformed LitSearch queries with a prompt-based classifier, using the query-labeling prompt.
This gives us an intent distribution over the adapted benchmark rather than a single undifferentiated query pool.
As shown in Figure~\ref{fig:query_type_distribution}, evidence-based queries dominate the paper-to-model recommendation setting, while the other categories appear much less frequently, but are still present.
We treat this analysis as a descriptive step before the main quantitative evaluation, since different intents may benefit differently from semantic anchoring and structure-aware retrieval.

\subsection{ModelCard-Level Quantitative Evaluation}
\label{sec:quantitative}

\begin{figure*}
\centering
\includegraphics[width=\linewidth]{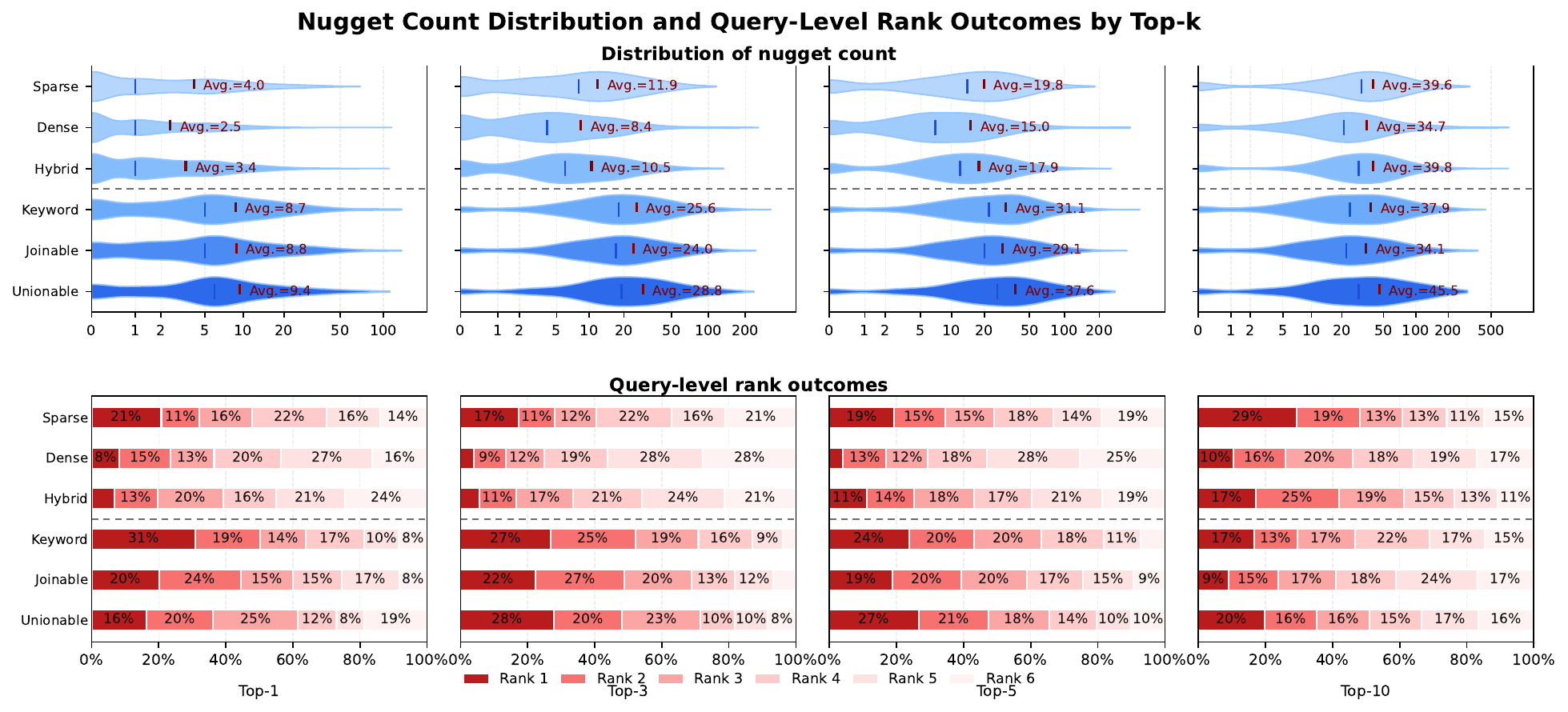}
\caption{Nugget coverage (top blue) and query-level rank (bottom red) share across retrieval methods under different top-$k$ budgets. 
The upper panel aggregates per-query nugget counts for each method, the dark blue vertical line shows the median, the red vertical line shows the average number of nuggets.  The bottom panel shows how often each method ranks first through sixth under the same budget. 
Across top-$1$, top-$3$, and top-$5$, unionable table search is the strongest structure-aware variants, though joinable and keyword are still out-competing most of the \baseline\ baselines where sparse retrieval is the strongest.
At top-$10$, the gap between methods becomes smaller and the relative advantage of the strongest structure-aware operator is less pronounced, indicating budget sensitivity.
}
\label{fig:result}
\end{figure*}

For each adapted query, we run all retrieval methods (three varients of \baseline\ and three varients of \oursearch) and collect the returned top-$k$ model-card candidates.
We then compute the nugget-count score defined in Section~\ref{sec:nugget-eval} for each returned set.
Because the same query can favor different evidence patterns, we report both the per-query distribution and the aggregate summary.
At the per-query level, the same nugget is counted only once even if it appears in multiple retrieved cards, so the score reflects distinct evidence rather than redundant copies of the same fact.
At the aggregate level, the mean coverage tells us which methods surface the broadest evidence mass across the full benchmark.

Note that even for top-1 queries that return a single model card, there may possibly be hundreds of unique nuggets satisfying a query.  As one example, for the query {\tt "Could you suggest models that investigate how many evidence sentences are needed for document-level RE?"}, the nugget creation recognizes that datasets are required to answer this query.  Furthermore, the model {\tt fblgit/juanako-7b-v1} has been run on dozens of benchmark datasets and report several performance metrics on each leading to 134 nuggets being created that match the query.  

To understand how retrieval strategy changes the models returned, we show two complementary views of the results in Figure~\ref{fig:result}.
The top (blue) panel is method-centered: it aggregates the number of nuggets satisfying the query over all adapted queries and shows the number of distinct nuggets  each retrieval method tends to surface, independent of any single query.
The 
bottom
panel is query-centered: for each query, we compare all methods and count how often each one lands at rank 1 through rank 6 under the same top-$k$ budget.
This split matters because the structure-aware family is not uniform.
Unionable, joinable, and keyword search impose different structural constraints, so their behavior depends on how directly the query aligns with the available table schemas and value patterns.
Likewise, the \baseline\ family is not interchangeable: sparse retrieval often remains competitive because exact lexical overlap can still capture strong task alignment in this benchmark.

The top panel illustrates  which methods surface the most nugget evidence on average, and the bottom panel illustrates which methods are most often near the top on a per-query basis.  
That second question is important because a method can have a strong average without being the most consistent winner, and vice versa.
Across top-$1$, top-$3$, and top-$5$, the overall pattern is stable: unionable is the strongest structure-aware operator, joinable is more selective and therefore succeeds less often, and keyword search is more brittle because it relies on overlapping vocabulary.
However, top-$10$ changes the picture: the ranking shifts enough that sparse and hybrid become more competitive, which shows that retrieval depth is not just a scaling detail, but a factor that influences accuracy.  

\subsection{Table-Level Qualitative Evaluation}
\label{sec:demo}

\begin{figure}[!h]
    \centering
    \includegraphics[width=\linewidth]{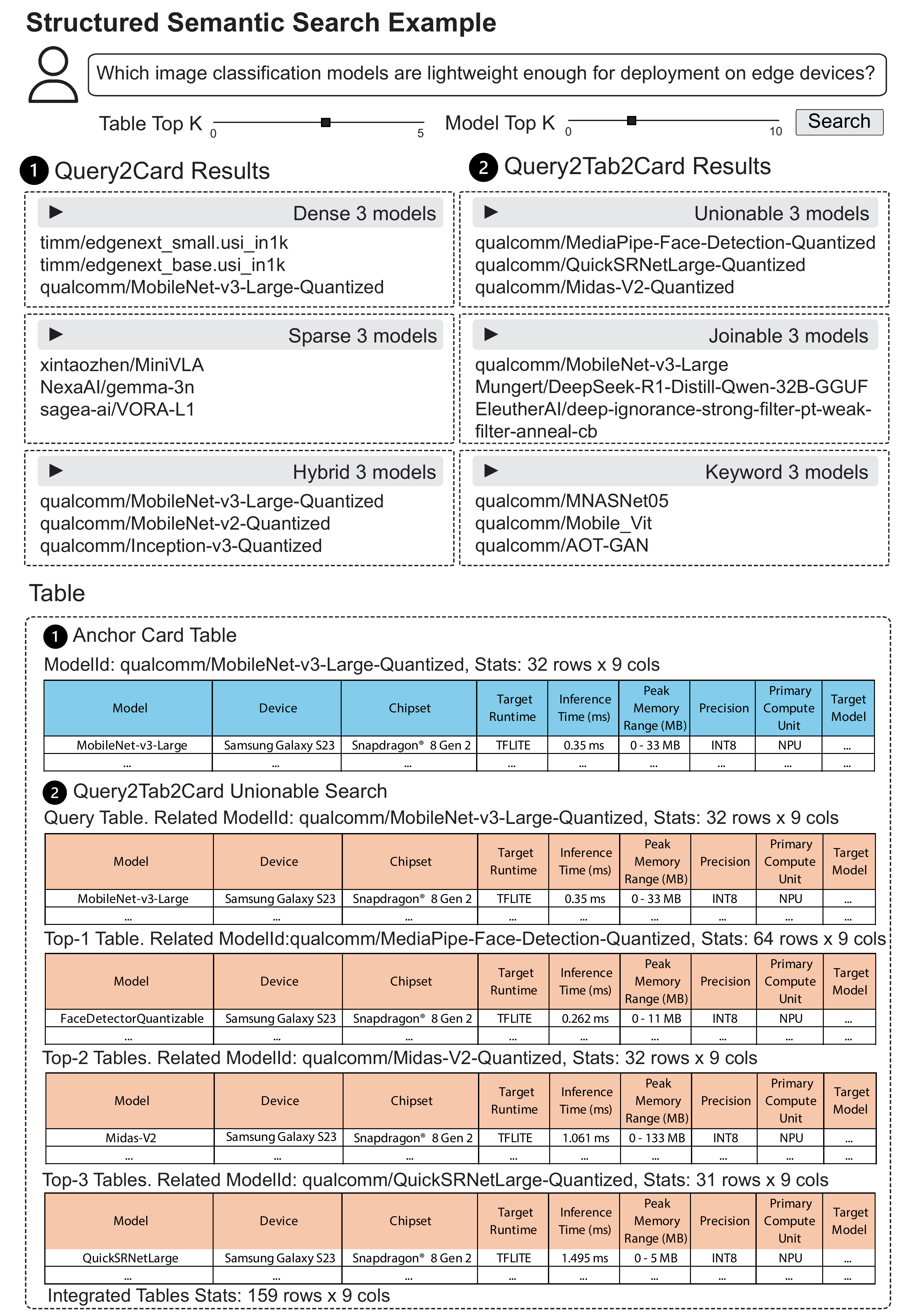}
    \caption{
(1) \baseline\ returns less diverse result sets of models than (2) \oursearch.
In addition, \oursearch\ expands a query-aligned seed table (from the anchor card) with related tables (we show the results of using unionability as the relatedness measure) from related models, enabling a broader and more transparent comparison across models.  The integration (union) of these tables also provides the user with  task-relevant information on the performance of related  models augmenting the card-search with valuable structured information.}
    \label{fig:benchmark_compare}
\end{figure}

We use a resource-constrained model selection query, {\tt "Which image classification models are lightweight enough for deployment on edge devices?"}, as a representative example in Figure~\ref{fig:benchmark_compare}.
The main insight is that the \baseline\ methods are generally returning less diverse sets of models but also return models whose cards may not contain interesting structured evidence useful to a data scientist in comparing the models.
Even when relevant cards are retrieved, users may still need to read long textual descriptions, and the information is often presented in inconsistent formats across cards, which makes quick comparison difficult.
In contrast, using \oursearch\ methods,  we ensurethe retrieved evidence is table-backed and thus already organized in a more consumable way. Furthermore, we present (and integrate) the tables used in our search as illustrated in Figure~\ref{fig:benchmark_compare}.   
In this example, this enables the system 
to produce a broad yet still coherent comparison view over attributes such as device, chipset, runtime, latency, memory usage, precision, and compute unit.
This is analogous to benchmark curation platforms such as Papers with Code\footnote{\url{https://paperswithcode.com}}, and suggests that table-centric retrieval can support the automatic construction of benchmark-style comparison tables by integrating compatible evidence from different model-card tables.

\begin{figure}[!h]
    \centering
    \includegraphics[width=\linewidth]{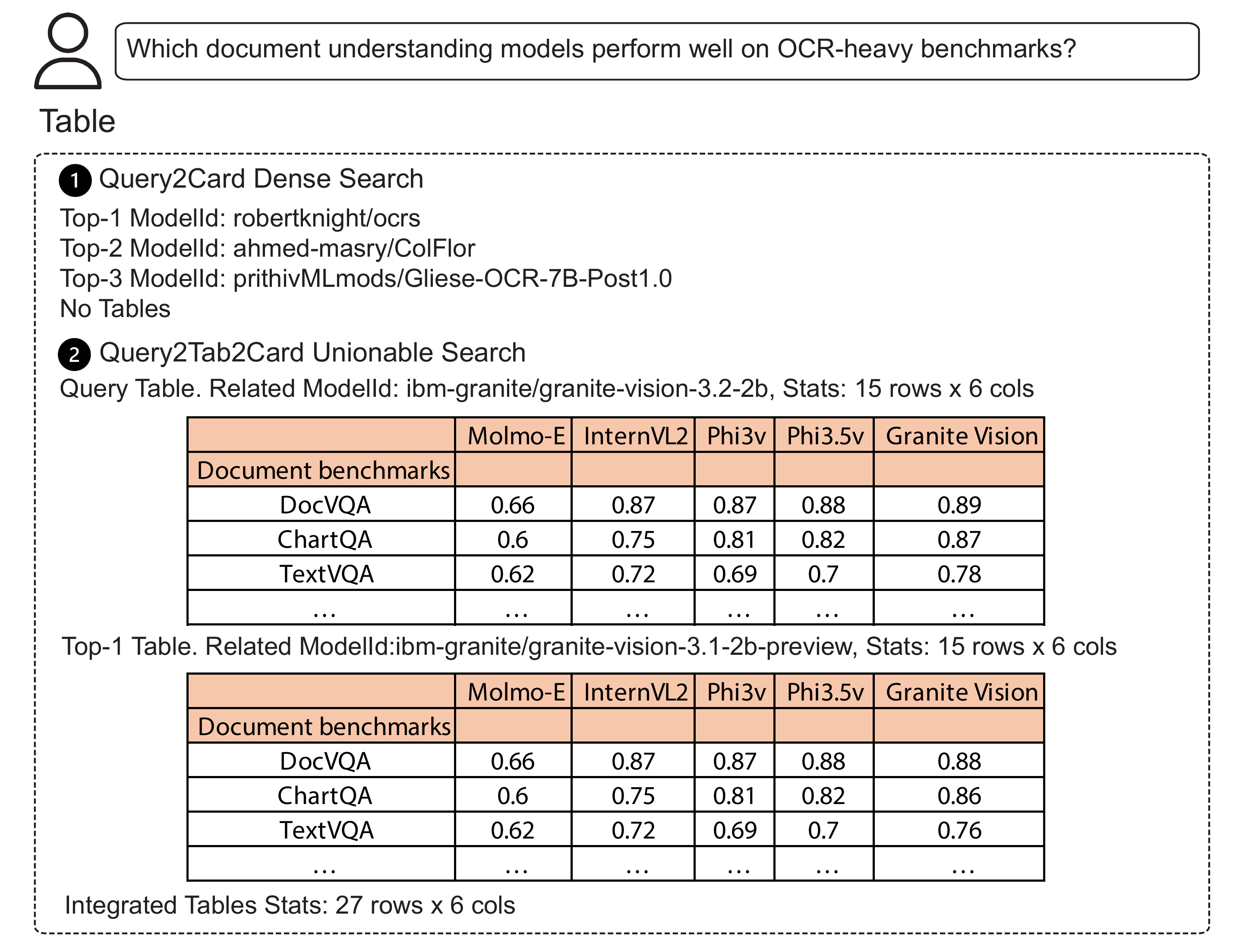}
    \caption{
    \baseline\ retrieves model cards without usable tables, whereas \oursearch\ returns benchmark tables aligned with the task, enabling direct comparison across models and supporting fine-grained version analysis within model family.}
    \label{fig:ocr_compare}
\end{figure}

Our second example considers a query on OCR-heavy document understanding models, where the goal is to compare benchmark performance across models.
As shown in Figure~\ref{fig:ocr_compare}, \oursearch\ retrieves a benchmark table whose columns are model names, including Molmo-E, InternVL2, Phi3V, Phi3.5V, and Granite Vision, while the first column lists document benchmarks.
This structure matches the task directly and supports table question answering over the retrieved evidence.

A key point is that dense model-level retrieval again fails to guarantee structured evidence: in this example, the top retrieved model cards contain no usable tables.
In contrast, \oursearch\ using unionable table search returns two tables with the same schema and nearly identical columns, differing mainly in one model column.
From the model identifiers, these tables appear to correspond to different versions within a model family, making this example directly relevant to benchmark comparison.

\section{Conclusion}
\label{sec:conclusion}

We presented \oursearch, a table-centric framework for model search in model lakes.
Rather than relying only on model-card-level semantic retrieval, \oursearch\ treats tables as searchable and integrable evidence units, which makes it possible to retrieve structured model information, surface more diverse evidence (and models), and assemble comparison-ready candidate sets.
Our evaluation uses 597 published LitSearch paper recommendation queries adapted to model recommendation, and our nugget-based quantitative evaluation measures how much query-relevant evidence each retrieval method surfaces at the model-card level.
A small qualitative case studies complement this  showing that integrated tables retrieved through the structure-aware pipeline are  coherent and comparison-ready than the unstructured evidence returned by semantic retrieval alone.
Taken together, the nugget-based quantitative results and the integrated table-based qualitative evidence support the central claim that table-centric retrieval is a useful complement to information-retrieval-based semantic model search for evidence-grounded decision making.

There are several directions for future work.
First, to integrate tables, we have used in this paper Alite~\cite{khatiwada2022integrating}, a scalable approach to integrating data lake tables that maximally integrates facts scattered across tuples in different tables.  However, Alite does not work well if one table is the transpose of another (or more generally if they exhibit schematic heterogeneity where data in one table is used as headers in another)~\cite{DBLP:conf/sigmod/Miller98}.  Such heterogeneity is common in model lakes~\cite{DBLP:journals/corr/abs-2512-16106} and more research is needed on how to best integrate these tables.
Second, many model cards remain incomplete or only partially table-backed, so future systems should infer or augment missing structure to improve search and integration over incomplete cards.
Third, the majority of our queries are evidence-based (Figure~\ref{fig:query_type_distribution}), it would be interesting to understand if the search performance trade-off change for workloads with different types of query intents.
Finally, we used nuggets comparatively, to compare different search strategies.  They have the potential to be used to create ground-truth for a model search benchmark, allowing for the reporting of precision and recall.

\begin{acks}
\end{acks}
\bibliographystyle{ACM-Reference-Format}
\bibliography{bib/topic-datalake,bib/topic-modellake,bib/topic-others,bib/topic-ourgroup, bib/topic-llm, bib/topic-scitable}



%
% ...

\end{document}